# Inferring scattering-type Scanning Near-Field Optical Microscopy Data from Atomic Force Microscopy Images

Stefan G. Stanciu*[1,2], Stefan R. Anton[1], Denis E. Tranca[1], George A. Stanciu[1], Bogdan Ionescu[2,3], Zeev Zalevsky[4], Binyamin Kusnetz[5,6], Jeremy Belhassen[5,6], Avi Karsenty*[5,6] and Gabriella Cincotti*[7]

[1]Center for Microscopy-Microanalysis and Information Processing, National University of Science and Technology POLITEHNICA Bucharest, 313 Splaiul Independentei, 060042, Romania

[2]CAMPUS Research Institute, National University of Science and Technology POLITEHNICA Bucharest, 313 Splaiul Independentei, 060042, Romania

[3]AI Multimedia Lab, National University of Science and Technology POLITEHNICA Bucharest, 313 Splaiul Independentei, 060042, Romania

[4]Faculty of Engineering and the Institute of Nanotechnology and Advanced Materials, Bar-Ilan University, Ramat Gan 5290002, Israel

[5]Advanced Laboratory of Electro-Optics (ALEO), Dept. of Applied Physic/Electro-Optics Engineering, Faculty of Engineering, Jerusalem College of Technology, 9116001, Jerusalem, Israel

[6]Nanotechnology Center for Research and Education, Jerusalem College of Technology, 9116001, Jerusalem, Israel

[7]Department of Civil, Computer Science and Aeronautical Technologies Engineering, University Roma Tre, Rome, Italy

***Corresponding authors:***
Stefan G. Stanciu: stefan.g.stanciu@upb.ro
Avi Karsenty: karsenty@jct.ac.il
Gabriella Cincotti: gabriella.cincotti@uniroma3.it

**ABSTRACT:** Optical nanoscopy is crucial in life and materials sciences, revealing subtle cellular processes and nanomaterial properties. Scattering-type Scanning Near-field Optical Microscopy (s-SNOM) provides nanoscale resolution, relying on the interactions taking place between a laser beam, a sharp tip and the sample. The Atomic Force Microscope (AFM) is a fundamental part of an s-SNOM system, providing the necessary probe-sample feedback mechanisms for data acquisition. In this Letter, we demonstrate that s-SNOM data can be partially inferred from AFM images. We first show that a generative artificial intelligence (AI) model (pix2pix) can generate synthetic s-SNOM data from experimental AFM images. Second, we demonstrate that virtual s-SNOM data can be extrapolated from knowledge of the tip position and, consequently, from AFM signals. To this end, we introduce an analytical model that explains the mechanisms underlying AFM-to-s-SNOM image translation. These insights have the potential to be integrated into future physics-informed explainable AI models. The two proposed approaches generate pseudo s-SNOM data without direct optical measurements, significantly expanding access to optical nanoscopy through widely available AFM systems. This advancement holds great promise for reducing both time and costs associated with nanoscale imaging.

Optical nanoscopy techniques are crucial in life and materials sciences[1, 2], enabling observations beyond the diffraction limit to study cellular processes and nanoscale material properties. Although fluorescence-based super-resolution microscopy routinely achieves resolutions surpassing the diffraction limit with one order of magnitude[3], with some variants being even capable of sub-nanometre resolutions [4, 5], techniques in this family require specialized probes, which limits their use in exploring the intrinsic physico-chemical properties of advanced nanomaterials and nanostructures, and of unlabelled biological specimens.

Tip-enhanced nanoscopy combines the high resolution of scanning probe microscopy with the sensitivity of optical techniques [7]. Using sharp AFM-like tips, methods such as Tip-Enhanced Raman Spectroscopy (TERS)[6], Tip-Enhanced Fluorescence (TEF) [7], Tip-Enhanced Photoluminescence (TEPL) [8], Tip-Enhanced Second-Harmonic Generation (SHG) [9, 10] and Photo-induced Force Microscopy (PiFM) [11] leverage tip-sample-light interactions for nanoscale optical characterization. These techniques achieve resolution limited not by wavelength, but by tip size, composition, and detector sensitivity—typically in the range of 5–30 nm, with some setups reaching sub-nanometre scales[8, 12].

Scattering-type Scanning Near-field Optical Microscopy (s-SNOM) is one of the most widely used tip-enhanced nanoscopy techniques, driving numerous discoveries[13-15]. It uses a sharp tip scanned over a sample and illuminated by a focused laser, generating a localized near field at the tip apex. This near-field interaction modulates the scattered light's amplitude and phase, depending on the local dielectric properties of the tip and sample[16]. s-SNOM enables quantitative nanoscale mapping of dielectric and optical properties[13, 17-19] and chemical composition [20], advancing research in condensed matter[21, 22], low-dimensional materials [23-25], and increasingly, biological systems[26-29].

At the core of any s-SNOM system lies an AFM system, enabling probe-sample feedback mechanisms required for image acquisition, and AFM images are typically recorded simultaneously with s-SNOM data, which is useful for placing the optical information into a topographic context. Interestingly, although AFM and s-SNOM techniques rely on completely different contrast mechanisms, with AFM being a non-optical technique that can probe with superb level of detail the sample's topography based on van-der-Waals interactions between the tip and the sample[30], and s-SNOM being an optical technique pooling from dielectric and absorption contrast[13], sample properties that account as the main contrast source in one of the two, exhibit also a parasitic influence on the second. For example, topographic elements are known to influence s-SNOM data[31-33], and the chemical composition of the sample can influence the topography read-out by AFM[34-36]. This bi-directional crosstalk makes the data collected by the two techniques correlated, to some extent.

Considering the relationship between AFM and s-SNOM data a natural question that arises is whether s-SNOM images can be inferred from AFM data. With respect to such applications, we recall that image-to-image translation applications in microscopy powered by Generative Adversarial Networks (GANs), represent a transformative advancement and paradigm shift in the field of microscopic imaging accelerating research, by making sample characterization more efficient and more accessible. For instance, GANs can provide facile virtual access to levels of resolution typically available in very expensive instruments, confined to a limited number of organizations worldwide[37, 38]. Additionally, GANs facilitate the translation of images between different modalities[39-41].

In this Letter, we present AFM to s-SNOM image translation results obtained with pix2pix[42], a GAN framework (**Fig. 1a**, adapted from Adıyaman et al.[43]) that was originally designed for image-to-image translation tasks for mass-consumption applications, such as converting sketches to photographs. In this proof-of-concept experiment, pix2pix was selected considering that its architecture is highly representative for this class of AI models. Pix2pix uses a U-Net as its generator architecture, creating an output image based on an input image while preserving both global and local image features through skip connections, while its patch-based discriminator evaluates the realism of small regions of the generated image. The training involves two loss functions: an adversarial loss to ensure realism and an L1 loss to preserve fidelity to the ground truth.

More specifically, we used the pix2pix framework to translate AFM images into s-SNOM amplitude images (**Fig. 1a**), which are generally acknowledged as a highly useful optical mean for identifying material contrasts and surface structures. As discussed in previous works, s-SNOM amplitude images provide information on the strength of the scattered near-field signal, being correlated with the local optical reflectivity or scattering efficiency of the sample[13]. Conversely, s-SNOM phase images, not addressed in this work, map the phase shift of the scattered signal relative to the incident light. s-SNOM phase images are sensitive to variations in composition, conductivity, and permittivity, and can differentiate materials with similar reflectivity but different optical properties[13, 17, 44].

Given that pseudo-heterodyne detection is the most common s-SNOM configuration—using a lock-in amplifier tuned to higher harmonics of the tip's tapping frequency combined with a reference mirror—we focused on generating synthetic s-SNOM images (2nd and 3rd harmonics: O2A and O3A) from the experimental AFM image. To obtain results for O2A and O3A s-SNOM signals, we constructed two training datasets, each composed of 1076 image pairs, acquired in diverse experiments on different materials, e.g.[18, 45, 46], as well as biological samples [47, 48], **Supplementary Fig. 1**. One dataset consists of AFM and s-SNOM O2A image pairs, and the other consists of AFM and s-SNOM O3A image pairs, acquired in both scanning directions (**Fig. 1.b**); AFM and s-SNOM are inherently registered upon data acquisition, as per standard s-SNOM imaging protocols. For training, each of the raw AFM and s-SNOM images was converted to PNG format and scaled down from 300 × 300 pixels to 256 × 256 pixels, to match the input requirements of the pix2pix model. The model was trained for 100 epochs with a learning rate of 0.0002, followed by another 100 epochs where the learning rate linearly decayed to 0. The batch size for training was 1 image, and the momentum term for Adam had a value of 0.5.

In **Fig. 1c)** we present side-by-side experimental and synthetic s-SNOM images obtained on a reference sample consisting of vanadium structures on a quartz substrate. While high-level features such as small details on the vanadium structure, or on the substrate are missing, its shape is correctly reproduced, and the signal intensity ratio between the quartz substrate and the vanadium structure on top are highly similar. In the bottom panel of **Fig. 1.c)** we present experimental/synthetic s-SNOM images on *Burkholderia cenocepacia* ATCC BAA-245 bacteria[47]. While the shape of the bacterial cells is retained in the synthetic s-SNOM images, in some cases small details such as the division septum, or border between two adjacent cells were not correctly reproduced. Furthermore, topographic artefacts, such as the yellow-coloured hallo surrounding some of the bacterial cells in the experimental s-SNOM imaging (that we associate to topography-induced artefacts), were not reproduced in the synthetic s-SNOM image, but similar yellow hallos were hallucinated in the pix2pix generated image for different cells. Next, in **Fig. 1d)** we provide two additional examples that allow more quantitative insights into AFM to s-SNOM pix2pix image translation, obtained for an AFM calibration grating (TGQ1, TipsNano) consisting of $SiO_2$ squares deposited on a Si substrate, also used in our previous works on s-SNOM[48]. While the shapes available in the experimental O2A and O3A s-SNOM images are overall well reproduced in the synthetic images produces by pix2pix, the results for O3A are better both in terms of shape fidelity, and pixel level signal variation, as observed in the profile lines. Additionally, the homogeneity of the synthetic signals obtained for the $SiO_2$ squares is better for the case of O3A. We highlight that the range of intensities of SiO and $SiO_2$ corresponding pixels is relatively similar between the compared profile lines, which we find to be an important output of the pix2pix model.

Overall, the preliminary results presented in **Fig. 1** are very promising, and they suggest that better trained or more refined AI models can potentially achieve AFM-to-s-SNOM image translation with close to full fidelity compared to experimental data. With respect to more sophisticated AI models, we recall that physics informed AI may represent an important avenue for the here addressed problem. This assumption is based, among others, on previous results on physics informed AI methods for s-SNOM, introduced for extracting quantitative dielectric permittivity data from raw s-SNOM data[49], corroborated with the overall success of physics-informed machine learning models[50]. To assist the future advent of physics-informed AI models for AFM to s-SNOM image to image translation, we introduce next an analytical model, that we entitled Vs-SNOM, that can be used to infer virtual s-SNOM amplitude data from AFM signals. The core concept of the model is presented next, and the schematic represented in **Fig. 2a**.

In a typical s-SNOM microscope, AFM and s-SNOM data are simultaneously registered during the sample scanning. For each position (*x,y*) of the tip (corresponding to an image pixel) the AFM amplitude $A(x,y)$, phase $\varphi(x,y)$ and topography $Z_S(x,y)$ signals are detected to generate the corresponding AFM images.

The time-dependent position of the tip depends on the AFM signals[51]:

$$z_{tip}(x,y,Z_S\,,t) = Z_S(x,y) + \frac{A(x,y)}{2}\{1 + sin[2\pi f_0 t + \varphi(x,y)]\} \qquad (1)$$

and on the oscillating frequency $f_0$. It is known that the probe deflects upon interactions between the tip and the sample due to attractive or repulsive forces: in our model, the nonlinear properties of the tip motion are implicitly expressed in the dependencies of the oscillation amplitude $A(x,y)$ and phase $\varphi(x,y)$ on the position (*x,y*)[52].

In the s-SNOM apparatus, a laser beam with wavelength $\lambda$ and average electric field $E_0$ is focused to the interaction region between the metallic tip and the sample. The light signal that reaches the photodetector is composed of four beams. The first term is the light scattered by the tip $E_1(x,y,Z_S,t)$, that is assumed to be a spherical wave generated from the dipole at the tip. The second signal is the light from the image dipole $E_2(x,y,Z_S,t)$ inside the sample, which is also modelled as a spherical wave. The third term is the light scattered by the sample $E_3$ (background field), that is assumed to be constant in time and space. Finally, there is the light reflected by the vibrating mirror $E_4(t)$ that oscillates with frequency *M.*

Many different approaches have been proposed in the literature to model the light scattered by the tip, assuming the scattering probe to be a polarizable sphere[16] or an ellipsoidal particle[53]. In our model, we assume that the light scattered by the tip is originated by a point-like source in ($x$, $y$, $z_{tip}$) at a distance $r$ (see Fig. 2a)

$$r = \sqrt{(x - D_x)^2 + \left(y - D_y\right)^2 + \left(z_{tip} - D_z\right)^2} \qquad (2)$$

from the detector located in ($D_x$, $D_y$, $D_z$), with

$$D = \sqrt{D_x^2 + D_y^2 + D_z^2}. \qquad (3)$$

Therefore, the light scatted by the tip can be written as

$$E_1(x,y,Z_S,t) = E_0 \frac{B}{r} exp\left(i \frac{2\pi}{\lambda} r\right) \qquad (4)$$

where $B$ is a suitable constant that takes the polarizability of the tip into account and $i$ is the imaginary unit. Note that the model can potentially be further extended using a more accurate description of the tip shape and polarizability[54]. In addition, we have assumed $E_1(x,y,Z_S,t)$ to be a spherical wave, but a more accurate laser model, for instance a Gaussian beam, can be used depending on the specifics of the experiment.

The tip dipole induces a charge distribution close to the sample surface that can be described as an image point-like dipole at

$$z'_{tip}(x,y,Z_S\,,t) = Z_S(x,y) - \frac{A(x,y)}{2}\{1 + sin[2\pi f_0 t + \varphi(x,y)]\} \quad (5)$$

that is the source for another spherical wave

$$E_2(x,y,Z_S,t) = E_0\beta(x,y)\frac{B}{r'}exp\left(i\frac{2\pi}{\lambda}r'\right), \quad (6)$$

where

$$r' = \sqrt{(x - D_x)^2 + \left(y - D_y\right)^2 + \left(z'_{tip} - D_z\right)^2} \quad (7)$$

is the distance of the image dipole from the detector and

$$\beta(x,y) = \frac{\varepsilon(x,y) - 1}{\varepsilon(x,y) + 1} \quad (8)$$

is a parameter that depends on the local sample relative permittivity $\varepsilon(x,y) = [n_e(x,y) + i\kappa(x,y)]^2$. $n_e(x,y)$ is the local refractive index (RI) and $\kappa(x,y)$ is the extinction coefficient. The complex refractive index $n_e(x,y) + i\kappa(x,y)$ is assumed to be constant with respect to $Z_s$, but the model could be further generalized considering a layered medium.

We observe that the light scatted by the tip $E_1(x,y,Z_S,t)$ is influenced by the field generated by the image dipole, and the parameter $B$ can be evaluated as[16]:

$$B = \frac{\alpha_0}{1 - \frac{\alpha_0\beta}{16\pi Z_S^3(x,y)}} \quad (9)$$

with

$$\alpha_0 = 4\pi\, R_0^3\,\frac{\varepsilon_{tip} - 1}{\varepsilon_{tip} + 2} \quad (10)$$

$R_0$ is the radius of the polarizable sphere (that is assumed *1* nm), $\varepsilon_{tip}$ the dielectric constant of the tip.

The background field $E_3$=$R$ $E_0$ has constant amplitude and phase (it is an unmodulated signal that does not depend on the tip position) and it is originated from the diffuse reflection at the surface; $R$ is the average reflection coefficient of the sample which we consider to be a real-valued, constant parameter for all tip positions ($x$,$y$), for the sake of simplicity. Also, the parameter $R$ is unknown in our experimental setup, but its value can be extracted by a calibration of the s-SNOM image of harmonic #0 (DC).

Therefore, the total scattered field can be expressed as

$$E_S(x,y,Z_S,t) = E_1(x,y,Z_S,t) + E_2(x,y,Z_S,t) + E_3 \qquad (11)$$
$$= E_0\left\{\frac{B}{r}exp\left(i\frac{2\pi}{\lambda}r\right) + \frac{B}{r'}\beta(x,y)exp\left(i\frac{2\pi}{\lambda}r'\right) + R\right\}$$

having substituted Eqs. (4) and (6) and the background field $E_3$=$R$ $E_0$.

In the far-field approximation, $r \sim D - \frac{(D_x x + D_y y + D_z z_{tip})}{D}$, and $r' \sim D - \frac{(D_x x + D_y y + D_z z'_{tip})}{D}$, and Eq. (11) becomes

$$E_S(x,y,Z_S,t) = E_0\frac{B}{D}exp\left[i\frac{2\pi}{\lambda}\left(D - \frac{D_x x + D_y y + D_z z_{tip}}{D}\right)\right] \qquad (1)$$
$$+ E_0\frac{B}{D}\beta(x,y)exp\left[i\frac{2\pi}{\lambda}\left(D - \frac{D_x x + D_y y + D_z z'_{tip}}{D}\right)\right]$$
$$+ E_0 R$$

Inserting Eqs. (1) and (5) into Eq. (12), we obtain:

$$E_S(x,y,Z_S,t) = E_0\frac{B}{D}exp\left[i\frac{2\pi}{\lambda}\left(D - \frac{D_x x + D_y y + D_z Z_S(x,y)}{D} - \frac{D_z A(x,y)\{1 + sin[2\pi f_0 t + \varphi(x,y)]\}}{2D}\right)\right] \qquad (13)$$
$$+E_0\frac{B}{D}\beta(x,y)exp\left[i\frac{2\pi}{\lambda}\left(D - \frac{D_x x + D_y y + D_z Z_S(x,y)}{D} + \frac{D_z A(x,y)\{1 + sin[2\pi f_0 t + \varphi(x,y)]\}}{2D}\right)\right] + E_0 R$$

The total scattered signal of Eq. (13) can be represented as a Fourier series with respect to the tip oscillation frequency $f_0$

$$E_S(x,y,Z_S,t) = E_0\sum_n \tau_n\,(x,y,Z_S)\exp(i2\pi n f_0 t) \qquad (2)$$

where the Fourier coefficients are

$$\tau_n(x,y,Z_S) = f_0\int_{-1/2f_0}^{1/2f_0} E_S(x,y,Z_S,t)\;exp(-i2\pi n f_0 t)dt \qquad (15)$$

$$n = 0, \pm 1, \pm 2, \ldots.$$

and they can be evaluated as

$$\tau_n(x,y,Z_S) = R\delta(n=0)$$
$$+\frac{B}{D}exp\left\{i\frac{2\pi}{\lambda}\left[D - \frac{D_x x + D_y y}{D} - \frac{D_z}{D}\left(Z_S + \frac{A(x,y)}{2}\right)\right] + in\varphi(x,y)\right\}J_{-n}\;\left[\frac{\pi}{\lambda}\frac{D_z}{D}A(x,y)\right]$$

$$+\frac{B}{D}\beta(x,y)exp\left\{i\frac{2\pi}{\lambda}\left[D-\frac{D_x x+D_y y}{D}-\frac{D_z}{D}\left(Z_S-\frac{A(x,y)}{2}\right)\right]+in\varphi(x,y)\right\}J_n\left[\frac{\pi}{\lambda}\frac{D_z}{D}A(x,y)\right]$$

(16)

where $\delta$ is the Dirac delta and $J_n$ is the Bessel function of first kind and order $n$

$$J_n(x)=\frac{1}{2\pi}\int_{-\pi}^{\pi}\exp[i\,(xsin\theta-n\theta)]d\theta \qquad (3)$$

Therefore, the modulus of the Fourier coefficient of the scattered signal of harmonic $n\geq 1$ is

$$|\tau_n(x,y,Z_S)|=\left|\frac{B}{D}J_n\left(\frac{\pi}{\lambda}\frac{D_z}{D}A(x,y)\right)\right|\left|\beta(x,y)exp\left[i\frac{2\pi}{\lambda}\frac{D_z}{D}A(x,y)\right]-1\right| \qquad n=\pm1,\pm3,..$$

$$|\tau_n(x,y,Z_S)|=\left|\frac{B}{D}J_n\left(\frac{\pi}{\lambda}\frac{D_z}{D}A(x,y)\right)\right|\left|\beta(x,y)exp\left[i\frac{2\pi}{\lambda}\frac{D_z}{D}A(x,y)\right]+1\right| \qquad n=\pm2,\pm4,\ldots \quad (18)$$

having used the property $J_n(x)=(-1)^n J_{-n}(x)$.

On the other hand, since the average reflection coefficient $R$ is much larger than the other terms in Eq. (16), for the harmonic *#0* we have

$$\begin{aligned}|\tau_0(x,y,Z_S)|^2 \cong\ & R^2+2R\left|\frac{B}{D}J_0\left[\frac{\pi}{\lambda}\frac{D_z}{D}A(x,y)\right]\right| \\ & \cdot cos\left\{\frac{2\pi}{\lambda}\left[D-\frac{D_x x+D_y y+D_z Z_S}{D}-\frac{D_z}{D}\frac{A(x,y)}{2}\right]\right\} \\ & +2R\left|\beta(x,y)\frac{B}{D}J_0\left[\frac{\pi}{\lambda}\frac{D_z}{D}A(x,y)\right]\right| \\ & \cdot cos\left\{\frac{2\pi}{\lambda}\left[D-\frac{D_x x+D_y y+D_z Z_S}{D}+\frac{D_z}{D}\frac{A(x,y)}{2}+\Psi(x,y)\right]\right\}\end{aligned} \qquad (19)$$

where $\Psi(x,y)$ is the phase of $\beta(x,y)$.

The near-field signal $E_S(x,y,Z_S,t)$ interferes with the reference beam $E_4(t)$ at the photodetector. Using a pseudo-heterodyne detection approach, the reference beam $E_4(t)$ is phase modulated by means of the mirror oscillation at frequency $M$. Therefore, the reference beam can also be expressed as a Fourier series

$$\begin{aligned}E_4(t)=\ & CE_0 exp\left[i\frac{2\pi}{\lambda}Wsin(2\pi Mt)+i\psi_R\right] \\ & =E_0\sum_m \rho_m\exp(i2\pi mMt)\end{aligned} \qquad (4)$$

where $W$ is the mirror oscillation amplitude, $C$ is a suitable complex constant and $\psi_R$ accounts for the average optical path difference between the signal and reference beam. The corresponding Fourier coefficients are

$$\rho_m=CJ_m\left(\frac{2\pi}{\lambda}W\right)exp(i\psi_R) \qquad (5)$$

Since the reference modulation frequency $M$ is much lower than the tip vibration frequency $f_0$, each of the scattered signal harmonics with frequency $nf_0$ splits into sidebands with frequencies $f_{n,m}=n f_0 + mM$ and the SNOM signal amplitude obtained by detector output demodulation at frequency $f_{n,m}$ is

$$S_{n,m}(x, y, Z_S) = |\rho_m||\tau_n(x, y, Z_S)| \quad (6)$$

The AFM parameters of Eq. (1) (i.e. $A(x,y)$ amplitude, $\varphi(x,y)$ phase and $Z_S(x,y)$ topography) are extracted from the AFM images (**Fig. 2b**) and used in the Vs-SNOM model. Please note that the topography signal has been recalibrated, as it is shown in **Fig. 2b**.

To facilitate implementation of the analytical model, and ensure experiment reproducibility, in the Supporting Information we provide the MATLAB script, *VIRTUAL_sSNOM.m*, which represents a basic level implementation of the Vs-SNOM analytical model. The script uses AFM data to infer virtual, pseudo-optical, s-SNOM data that partially reproduces experimental s-SNOM images in content and resolution. It has been elaborated based on s-SNOM and AFM data sets collected with a NeaSNOM system (NeaSpec, Germany), and hence it refers to file naming conventions proprietary to this system. Information on these conventions, together with exemplary raw NeaSNOM AFM and s-SNOM datasets, were previously provided in our past work introducing the curated SSNOMBACTER dataset[47]. The script uses as a set of Portable Network Graphics (**.png)* files, converted from raw s-SNOM data files, and provides as output a series of *.png* files presented in Supplementary Table 1.

As a proof-of-concept, we provide an example obtained with *VIRTUAL_sSNOM.m* script on the TGQ1 (TipsNano) sample with $SiO_2$ squares deposited on a Si substrate, discussed also in **Fig. 1d)**. The AFM and s-SNOM images were acquired on this sample using a Mikromasch Hq:NSC19/Cr-Au gold coated probe with $a$ < 35nm tip radius, resonance frequency 65 KHz and force constant 0.5 N/m that was kept in an oscillating mode at frequency $f_0$ = 79 KHz when imaging calibration samples. Other measurement parameters for this example are provided in Supplementary Table 2.

It should be considered that the position of the detector ($D_x$, $D_y$, $D_z$), which is an important parameter in our model, (see **Fig. 2a**), was not precisely measured in our experimental setup, so an assumption was made in the MATLAB script, using values that fit better with the measured SNOM data. Furthermore, in the script, we assumed $m$ = 0, for sake of simplicity, and using Eqs. (18) and (19), we have numerically evaluated $|\tau_0(x, y, Z_S)|$, $|\tau_1(x, y, Z_S)|$, $|\tau_2(x, y, Z_S)|$, and $|\tau_3(x, y, Z_S)|$ signals that correspond to the SNOM amplitude signals corresponding to detection on DC, 1st harmonic (tapping frequency of the tip), 2nd and 3rd harmonics.

In **Fig. 2c** we present the outputs of our MATLAB script, consisting in experimentally measured s-SNOM and computationally calculated Vs-SNOM images, together with profile lines plotted across these to show similarity, for four cases: DC, 1st harmonic (the tapping frequency of the tip), 2nd harmonic, and 3rd harmonic. The experimental images for DC acquisition, where no modulation of the optical signals occurs, exhibit significant noise and signal inhomogeneities. The above introduced model is unable to reproduce noise, thus the experimental and virtual s-SNOM images are significantly different. Yet, the model is enough accurate to reproduce the interference fringes of Eq. (19). For the s-SNOM images where detection is performed at harmonics ($1^{st}$, $2^{nd}$ and $3^{rd}$) of the tapping frequency of the tip, the Vs-SNOM images share significant similarities with the experimental ones, which can be observed both at qualitative levels in the images displayed in **Fig 2c**, and quantitatively, with the help of the plotted profile lines. Considering these, we argue that the proposed model can represent a useful starting point for developing physics-informed AI models for synthetic s-SNOM imaging, where the outputs of a generative model are filtered through a mathematical formalism that explains the interdependencies of AFM and s-SNOM images.

In summary, we present two complementary methods for virtually generating pseudo-s-SNOM images from AFM data without a physical s-SNOM system. The first uses an off-the-shelf AI model (pix2pix) trained on paired AFM and s-SNOM images for image translation. The second is a physics-based approach that estimates s-SNOM images by modelling correlations between AFM and s-SNOM data, that depend on a series of acquisition parameters. While preliminary, these two methods show promise for evolving into a physics-informed AI model and could extend to other tip-enhanced techniques (e.g., TEF, TEPL, PIFM, AFM-IR). This approach is especially valuable given the accessibility of AFM compared to the cost and complexity of s-SNOM.

**Acknowledgments:** SGS, DET, and GAS acknowledge the financial support of UEFISCDI Grants PN-IV-P7-7.1-PED-2024-2374 (POLYNANO) and PN-IV-P8-8.3-PM-RO-TR-2024-0068 (CONAGAI).

## Figures

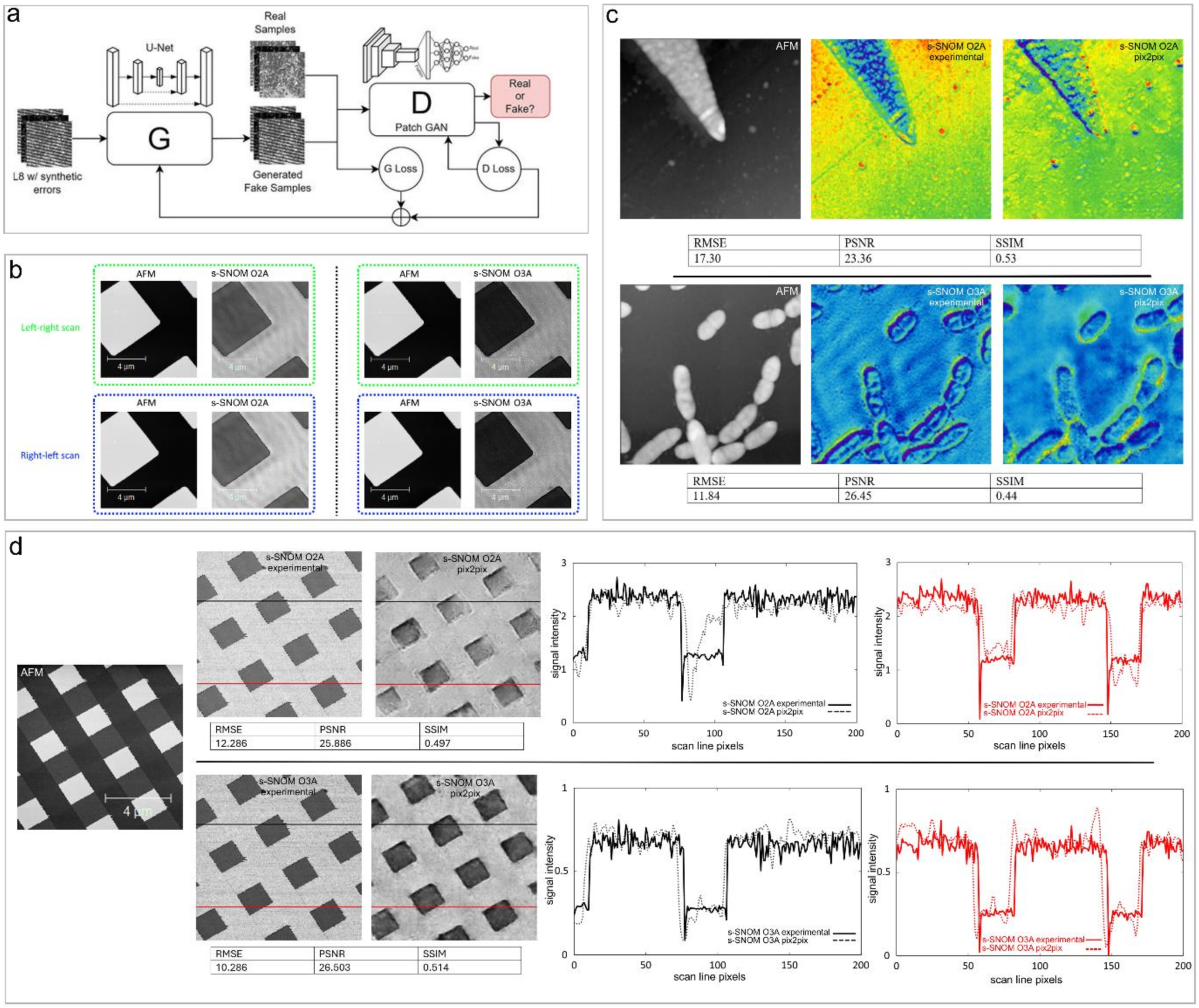

**Fig. 1.** AFM-to-s-SNOM amplitude image translation using the pix2pix GAN model. A) Schematic of the pix2pix architecture (adapted[43] under CC-BY 4.0 license terms). B) Example pairs from the training dataset. C) Proof-of-concept image translations: top—vanadium structures on a quartz substrate; bottom—prokaryotic cell samples, Burkholderia cenocepacia ATCC BAA-245 bacteria. Underneath the images we present results of three image quality metrics, Root Mean Square Error (RMSE), Peak Signal to Noise Ratio (PSNR) and Structural Similarity Index (SSIM). D) Case study: AFM-to-s-SNOM translation for an TGQ1 (TipsNano) calibration grating with $SiO_2$ squares on a Si substrate.

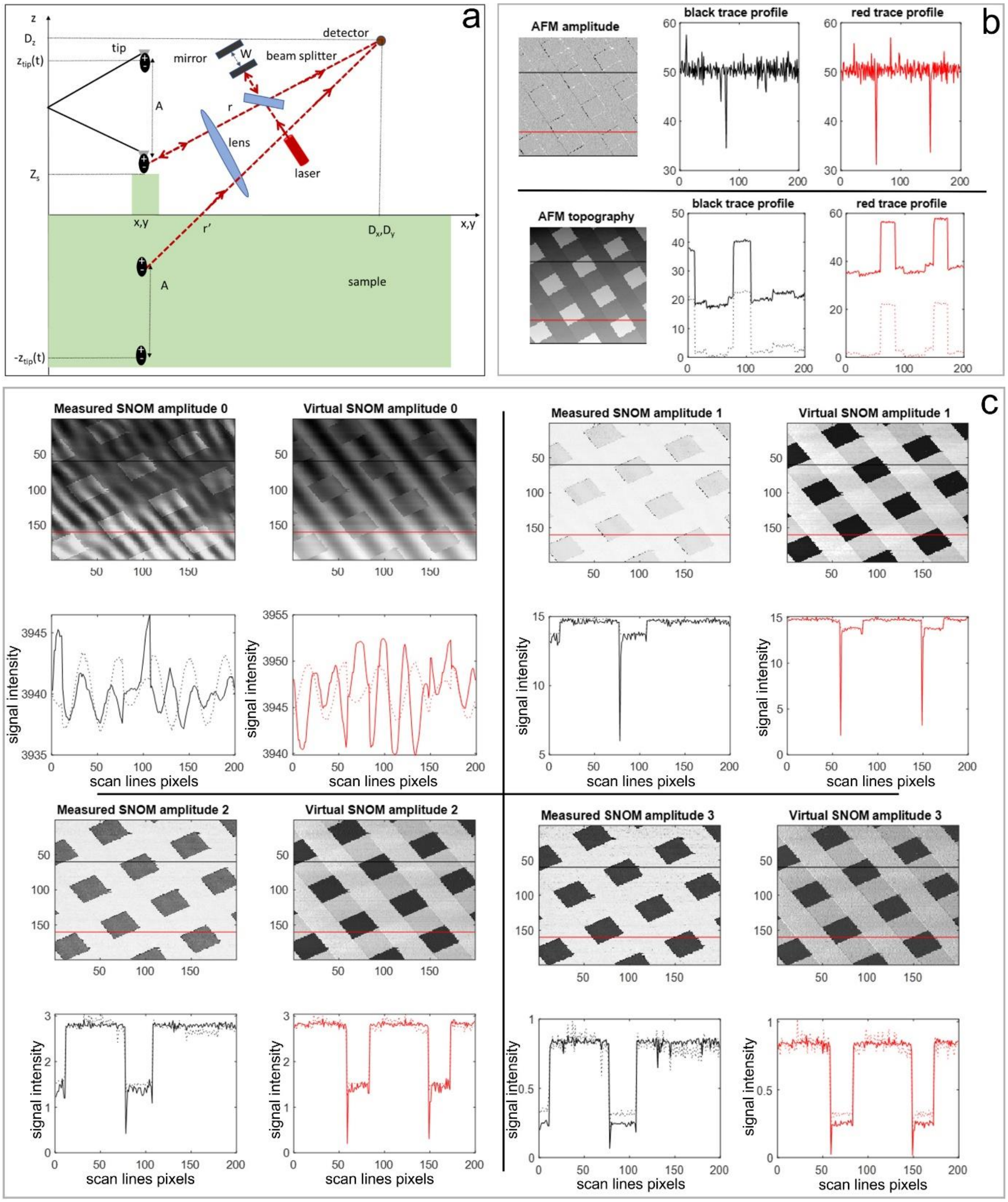

**Fig. 2.** Vs-SNOM concepts and results. a) Schematic representation of the Vs-SNOM analytical model. The s-SNOM scattered field is simulated using knowledge of the tip's position $z_{tip}(t)$ (from the AFM data) and the refractive index of the sample. The tip is represented by a dipole, oscillating along the $z$ axis (Eq. (11)); b) Images and trace profiles of the AFM amplitude *A(x,y)* and topography $Z_S(x,y)$ signals (before and after recalibration). (top) AFM amplitude *A(x,y)* signal. Two AFM amplitude signals *A(x,y)* measured along the black and red traces. (bottom)

AFM topography $Z_S(x,y)$ signal. Two AFM topography $Z_S(x,y)$ signals measured along the black and red traces (full line: measured signal, dotted line: recalibrated); c) Comparison of experimental and analytical SNOM results using the Vs-SNOM model. Solid lines represent measured data, and dotted lines represent virtual s-SNOM simulations. Measured and virtual SNOM amplitude at: (top left) DC; (top right) 1$^{st}$ harmonic (tip tapping frequency); (bottom left) 2$^{nd}$ harmonic; (bottom right) 3$^{rd}$ harmonic.

# Inferring scattering-type Scanning Near-Field Optical Microscopy Data from Atomic Force Microscopy Images

## Supporting Information

Stefan G. Stanciu*[1.2], Stefan R. Anton[1], Denis E. Tranca[1], George A. Stanciu[1], Bogdan Ionescu[2,3], Zeev Zalevsky[4], Binyamin Kusnetz[5,6], Jeremy Belhassen[5,6], Avi Karsenty*[5,6] and Gabriella Cincotti*[7]

[1]Center for Microscopy-Microanalysis and Information Processing, National University of Science and Technology POLITEHNICA Bucharest, 313 Splaiul Independentei, 060042, Romania

[2]CAMPUS Research Institute, National University of Science and Technology POLITEHNICA Bucharest, 313 Splaiul Independentei, 060042, Romania

[3]AI Multimedia Lab, National University of Science and Technology POLITEHNICA Bucharest, 313 Splaiul Independentei, 060042, Romania

[4]Faculty of Engineering and the Institute of Nanotechnology and Advanced Materials, Bar-Ilan University, Ramat Gan 5290002, Israel

[5]Advanced Laboratory of Electro-Optics (ALEO), Dept. of Applied Physic/Electro-Optics Engineering, Faculty of Engineering, Jerusalem College of Technology, 9116001, Jerusalem, Israel

[6]Nanotechnology Center for Research and Education, Jerusalem College of Technology, 9116001, Jerusalem, Israel

[7]Department of Civil, Computer Science and Aeronautical Technologies Engineering, University Roma Tre, Rome, Italy

Corresponding authors:

Stefan G. Stanciu: stefan.g.stanciu@upb.ro
Avi Karsenty: karsenty@jct.ac.il
Gabriella Cincotti: gabriella.cincotti@uniroma3.it

## List of Supporting Information

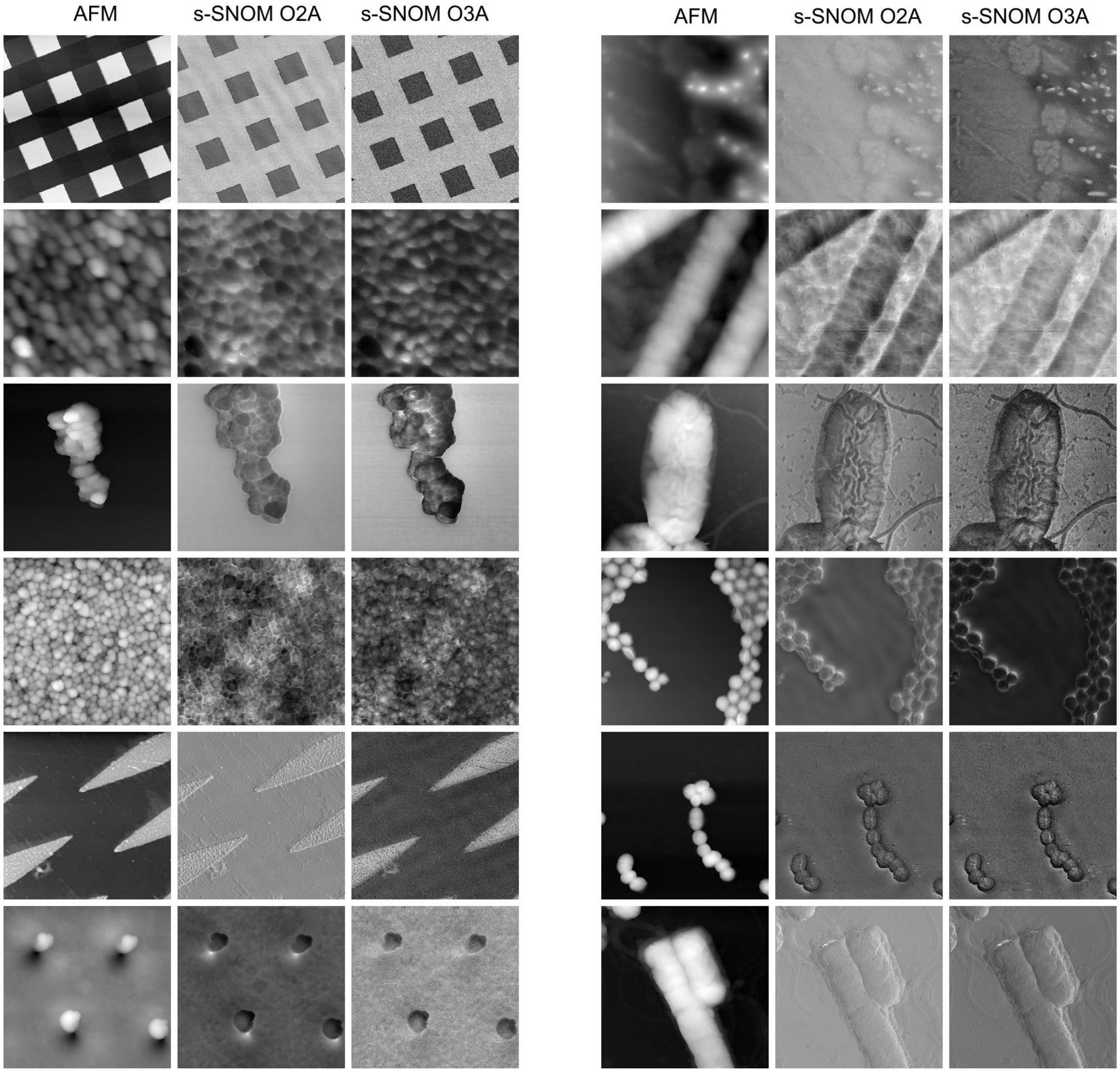

**Supplementary Fig. 1.** Examples from the training data set comprised of AFM, s-SNOM O2A and s-SNOM O3A images acquired on various materials and biological samples. **Left:** (from top to bottom): AFM calibration grating (TGQ1, TipsNano) consisting of $SiO_2$ squares deposited on a Si substrate (imaged by s-SNOM also in previous works[1]); nanoporous copper[2]; mmicroporous platinum-lead alloy; nanoporous silver[3]; vanadium structures on quartz substrate; poly(methyl methacrylate) (PMMA) modified by tip-enhanced fs-laser illumination[4]. **Right:** zebrafish retina[1, 5]; collagen; *Pseudomonas aeruginosa* bacterial cells[6]; *Staphylococcus auerus* bacterial cells[6]; *Enterococcus faecalis* bacterial cells[6]; *Bacillus subtilis subs. Spizizenii* bacterial cells[6].

| **Result file** | **Definition** |
|---|---|
| 1112 PH Si-SiO2_3 M1A.png | AFM amplitude $A(x,y)$ |
| 1112 PH Si-SiO2_3 M1P.png | AFM phase $\varphi(x,y)$ |
| 1112 PH Si-SiO2_3 Z.png | AFM Topography $Z_S(x,y)$ |
| 1112 PH Si-SiO2_3 O 0 A.png | SNOM amplitude harmonic #0 |
| 1112 PH Si-SiO2_3 O 5 A.png | SNOM amplitude harmonic #5 |
| 1112 PH Si-SiO2_3 O 0 P.png | SNOM phase harmonic #0 |
| 1112 PH Si-SiO2_3 O 5 P.png | SNOM phase harmonic #5 |

**Supplementary Table 1.** Filenames and meaning of the images resulted from applying the Vs-SNOM.m script.

| **Parameter** | **Values** | **Units** |
|---|---|---|
| Scan | 2D | |
| Layers | Si, $SiO_2$ | |
| Scanner Center Position (X, Y) | 50.00, 50.00 | [µm] |
| Rotation | 0 | [°] |
| Scan Area (X, Y, Z) | 10.000, 10.000, 0.000 | [µm] |
| Pixel Area (X, Y, Z) | 200, 200, 1 | [px] |
| Averaging | 1 | |
| Integration time | 3 | [ms] |
| NIR Laser Source | 1550 | [nm] |
| Target Wavelength | | [µm] |
| Demodulation Mode | PsHet | |
| Tip Frequency | 79,040.996 | [Hz] |
| Tip Amplitude | 454.700 | [mV] |
| Tapping Amplitude | 51.171 | [nm] |
| Modulation Frequency | 290.592 | [Hz] |
| Modulation Amplitude | 14,400.000 | [mV] |
| Modulation Offset | -2,386.496 | [mV] |
| Setpoint | 80.23 | [%] |
| Regulator (P, I, D) | 4.236081, 8.641345, 1.000000 | |
| Tip Potential | 0.000 | [mV] |
| M1A Scaling | 9.041 | [nm/V] |

**Supplementary Table 2.** s-SNOM data acquisition parameters used in the Matlab implementation example of the Vs-SNOM model.

*VIRTUAL_sSNOM.m* Matlab (The MathWorks, Inc, USA) script.

```
clear all
close all
lambda=1550; %wavelength nm

% range of data taken from Gwyddion
% % the  AFM and SNOM images have been previously processed using Gwyddion
% to extract the actual values of all the AFM and SNOM measured data.

AFMPhase_min=1; %
AFMPhase_max=2.1; % I guess they are radiants
AFMTopography_max=64.9;
AFMTopography_min=0; %
AFMAmplitude_min=23.5; % nm
AFMAmplitude_max=60.6;
SNOM0min=3935;
SNOM0max=3954;
SNOM1min=0.6;
SNOM1max=15.4;
SNOM2min=0.1;
SNOM2max=3;
SNOM3min=0.01;
SNOM3max=0.91;
SNOM4min=0.01;
SNOM4max=0.32;
SNOM5min=0.01;
SNOM5max=0.13;


%READ IMAGES
imageAFMAmplitude=rgb2gray(imread(char("2019-02-27 1112 PH Si-SiO2_3"+" M1A.png"))); % AFM amplitude
imageAFMPhase=rgb2gray(imread(char("2019-02-27 1112 PH Si-SiO2_3"+" M1P.png"))); % AFM phase
imageAFMTopography=rgb2gray(imread(char("2019-02-27 1112 PH Si-SiO2_3"+" Z.png"))); % Topography

for j=1:6
    numero=j-1;
imageSNOMAmplitude(:,:,j)=rgb2gray(imread(char("2019-02-27 1112 PH Si-SiO2_3"+" O"+numero+"A.png")));
% s-SNOM amplitude
imageSNOMPhase(:,:,j)=rgb2gray(imread(char("2019-02-27 1112 PH Si-SiO2_3"+" O"+numero+"P.png"))); % s-
SNOM phase
end


% area 10x10 um
start=1;
dimension_pixel=10000/200;% nm
%position of the detector (an assumption was made by numerical fitting)

Dx=0.8; % assumptions were made on the position of the detector
Dy=-0.48;
Dz=7.9;
```

```
% The script also two scans (plotted with black and red color) at two different
% heights in the image
level=60; % level of the traces are plotted
level2=160; % level of the traces are plotted

%starting point
start_x=1;
start_y=1;

dimension_y=size(imageAFMAmplitude,1)-1;
dimension_x=size(imageAFMAmplitude,2)-1;
[x,y]=meshgrid(start_x:dimension_x,start_y:dimension_y);
xx=linspace(start_x,dimension_x,dimension_x-start_x+1);
yy=linspace(start_y,dimension_y,dimension_y-start_y+1);

% rescaling the data extracted by the .png images using the actual values of the parameters (taken using Gwyddion)
% 8-bit images so max=255 e min=0
MAXI=255;
MINI=0;

AFMAmplitude=double(imageAFMAmplitude(start_y:dimension_y,start_x:dimension_x));
AFMAmplitude=(AFMAmplitude-MINI)*(AFMAmplitude_max-AFMAmplitude_min)/(MAXI-
MINI)+AFMAmplitude_min;

AFMPhase=double(imageAFMPhase(start_y:dimension_y,start_x:dimension_x));
AFMPhase=(AFMPhase-MINI)*(AFMPhase_max-AFMPhase_min)/(MAXI-MINI)+AFMPhase_min;

AFMTopography=double(imageAFMTopography(start_y:dimension_y,start_x:dimension_x));
AFMTopography=(AFMTopography-MINI)*(AFMTopography_max-AFMTopography_min)/(MAXI-
MINI)+AFMTopography_min;

for j=1:6
SNOMAmplitude(:,:,j)=double(imageSNOMAmplitude(start_y:dimension_y,start_x:dimension_x,j));
SNOMFase(:,:,j)=double(imageSNOMPhase(start_y:dimension_y,start_x:dimension_x,j));
end


% Rescaling
SNOM0=(SNOMAmplitude(:,:,1)-min(min(SNOMAmplitude(:,:,1))))*(SNOM0max-
SNOM0min)/(max(max(SNOMAmplitude(:,:,1)))-min(min(SNOMAmplitude(:,:,1))))+SNOM0min;
SNOM1=(SNOMAmplitude(:,:,2)-min(min(SNOMAmplitude(:,:,2))))*(SNOM1max-
SNOM1min)/(max(max(SNOMAmplitude(:,:,2)))-min(min(SNOMAmplitude(:,:,2))))+SNOM1min;
SNOM2=(SNOMAmplitude(:,:,3)-min(min(SNOMAmplitude(:,:,3))))*(SNOM2max-
SNOM2min)/(max(max(SNOMAmplitude(:,:,3)))-min(min(SNOMAmplitude(:,:,3))))+SNOM2min;
SNOM3=(SNOMAmplitude(:,:,4)-min(min(SNOMAmplitude(:,:,4))))*(SNOM3max-
SNOM3min)/(max(max(SNOMAmplitude(:,:,4)))-min(min(SNOMAmplitude(:,:,4))))+SNOM3min;
SNOM4=(SNOMAmplitude(:,:,5)-min(min(SNOMAmplitude(:,:,5))))*(SNOM4max-
SNOM4min)/(max(max(SNOMAmplitude(:,:,5)))-min(min(SNOMAmplitude(:,:,5))))+SNOM4min;
SNOM5=(SNOMAmplitude(:,:,6)-min(min(SNOMAmplitude(:,:,6))))*(SNOM5max-
SNOM5min)/(max(max(SNOMAmplitude(:,:,6)))-min(min(SNOMAmplitude(:,:,6))))+SNOM5min;

% the position of the tip is NOT the Topography signal
% it should be recalibrated
%
ztip=AFMTopography;
```

```
    zup=ztip;
zup(zup == 0 ) = NaN;
sub=zeros(dimension_y-start_y+1,1);
sub=min(zup,[],2);
sub2=xx*2/199;
ztip=ztip-sub-sub2;
ztip(ztip <0) = 0;

% dielectric constant of SiO2 3.9, of Si=11.7;
beta=(3.9-1)/(3.9+1);

radius=1; %nm (this is empirical, the tip is 35 nm long, and the radius of the dipole is assumed 1 nm)

alfa0=4*pi*radius^3*(11-1)/(11+2)/3;
alfa=alfa0./(1-alfa0*beta./(16*pi*(ztip+radius).^3));

SNOM0virtual=mean(SNOM0,2)+abs(alfa.*besselj(0,pi*Dz*AFMAmplitude/lambda)).*(cos(2*pi*((Dx*x+Dy*y)*
dimension_pixel+Dz*(ztip+AFMAmplitude/2))/lambda)+beta.*cos(2*pi*((Dx*x+Dy*y)*dimension_pixel-
Dz*(ztip+AFMAmplitude/2))/lambda));
SNOM1virtual=13.1+1.27*abs(alfa.*besselj(1,pi*Dz*AFMAmplitude/lambda)).*sqrt(1+beta.^2+2*beta*cos(4*pi*
Dz*(ztip+AFMAmplitude/2)/lambda)); % the constants before the formula have been introduced to match the average
value and range of measured data
SNOM2virtual=0.7+7.8*abs(alfa.*besselj(2,pi*Dz*AFMAmplitude/lambda)).*sqrt(1+beta.^2+2*beta*cos(4*pi*Dz
*(ztip+AFMAmplitude/2)/lambda)); % the constants before the formula have been introduced to match the average
value and range of measured data
SNOM3virtual=22.57*abs(alfa.*besselj(3,pi*Dz*AFMAmplitude/lambda)).*sqrt(1+beta.^2+2*beta*cos(4*pi*Dz*(
ztip+AFMAmplitude/2)/lambda)); % the constants before the formula have been introduced to match the average
value and range of measured data

figure (1)
subplot(2,3,1),imshow(imageAFMAmplitude)
hold on,plot(xx,level*ones(1,dimension_x-start_x+1), '-k') %plots the black trace
plot(xx,level2*ones(1,dimension_x-start_x+1), '-r') %plots the red trace
title('AFM amplitude')
subplot(2,3,2),plot(xx,AFMAmplitude(level,:),'-k')
title('black trace profile')
subplot(2,3,3),plot(xx,AFMAmplitude(level2,:), '-r')
title('red trace profile')
subplot(2,3,4),imshow(imageAFMTopography)
hold on,plot(xx,level*ones(1,dimension_x-start_x+1), '-k') %plot the line black
plot(xx,level2*ones(1,dimension_x-start_x+1), '-r') %plot the line red
title('AFM topography')
subplot(2,3,5),plot(xx,ztip(level,:),':k')
hold on
plot(xx,AFMTopography(level,:),'-k')
title('black trace profile')
subplot(2,3,6),plot(xx,ztip(level2,:), ':r')
hold on
plot(xx,AFMTopography(level2,:),'-r')
title('red trace profile')

figure (2)
subplot(2,2,1),image(xx,yy,(SNOM0-SNOM0min)/(SNOM0max-SNOM0min),'CDataMapping','scaled')
title('Measured SNOM amplitude 0')
hold on, plot(xx,level*ones(1,dimension_x-start_x+1), '-k') %plot the line black
plot(xx,level2*ones(1,dimension_x-start_x+1), '-r') %plot the line red
```

```
subplot(2,2,2),image(xx,yy,(SNOM0virtual-SNOM0min)/(SNOM0max-SNOM0min),'CDataMapping','scaled')
title('Virtual SNOM amplitude 0 ')
hold on,plot(xx,level*ones(1,dimension_x-start_x+1), '-k') %plot the line black
plot(xx,level2*ones(1,dimension_x-start_x+1), '-r') %plot the line red
colormap gray
subplot(2,2,3),plot(xx,SNOM0(level,:),'-k')
hold on,plot(xx,SNOM0virtual(level,:),':k')
subplot(2,2,4),plot(xx,SNOM0(level2,:),'-r')
hold on,plot(xx, SNOM0virtual(level2,:),':r')

figure (3)
subplot(2,2,1),image(xx,yy,(SNOM1-SNOM1min)/(SNOM1max-SNOM1min),'CDataMapping','scaled')
title('Measured SNOM amplitude 1')
hold on,plot(xx,level*ones(1,dimension_x-start_x+1), '-k') %plot the line black
plot(xx,level2*ones(1,dimension_x-start_x+1), '-r') %plot the line red
subplot(2,2,2),image(xx,yy,(SNOM1virtual-SNOM1min)/(SNOM1max-SNOM1min),'CDataMapping','scaled')
title('Virtual SNOM amplitude 1')
hold on,plot(xx,level*ones(1,dimension_x-start_x+1), '-k') %plot the line black
plot(xx,level2*ones(1,dimension_x-start_x+1), '-r') %plot the line red
colormap gray
subplot(2,2,3),plot(xx,SNOM1(level,:),'-k')
hold on,plot(xx,SNOM1virtual(level,:),':k')
subplot(2,2,4),plot(xx, SNOM1(level2,:),'-r')
hold on,plot(xx,SNOM1virtual(level2,:),':r')


figure (4)
subplot(2,2,1),image(xx,yy,(SNOM2-SNOM2min)/(SNOM2max-SNOM2min),'CDataMapping','scaled')
title('Measured SNOM amplitude 2')
hold on,plot(xx,level*ones(1,dimension_x-start_x+1), '-k') %plot the line black
plot(xx,level2*ones(1,dimension_x-start_x+1), '-r') %plot the line red
subplot(2,2,2),image(xx,yy,(SNOM2virtual-SNOM2min)/(SNOM2max-SNOM2min),'CDataMapping','scaled')
title('Virtual SNOM amplitude 2')
hold on,plot(xx,level*ones(1,dimension_x-start_x+1), '-k') %plot the line black
plot(xx,level2*ones(1,dimension_x-start_x+1), '-r') %plot the line red
colormap gray
subplot(2,2,3),plot(xx,SNOM2(level,:),'-k')
hold on,plot(xx,SNOM2virtual(level,:),':k')
subplot(2,2,4),plot(xx, SNOM2(level2,:),'-r')
hold on,plot(xx,SNOM2virtual(level2,:),':r')

figure (5)
subplot(2,2,1),image(xx,yy,(SNOM3-SNOM3min)/(SNOM3max-SNOM3min),'CDataMapping','scaled')
title('Measured SNOM amplitude 3')
hold on,plot(xx,level*ones(1,dimension_x-start_x+1), '-k') %plot the line black
plot(xx,level2*ones(1,dimension_x-start_x+1), '-r') %plot the line red
subplot(2,2,2),image(xx,yy,(SNOM3virtual-SNOM3min)/(SNOM3max-SNOM3min),'CDataMapping','scaled')
title('Virtual SNOM amplitude 3')
hold on,plot(xx,level*ones(1,dimension_x-start_x+1), '-k') %plot the line black
plot(xx,level2*ones(1,dimension_x-start_x+1), '-r') %plot the line red
colormap gray
subplot(2,2,3),plot(xx,SNOM3(level,:),'-k')
hold on,plot(xx,SNOM3virtual(level,:),':k')
subplot(2,2,4),plot(xx, SNOM3(level2,:),'-r')
hold on,plot(xx,SNOM3virtual(level2,:),':r')
```